\documentclass[3p,times,procedia]{elsarticle}
\usepackage{nupha_ecrc}

\volume{00}

\firstpage{1}

\journalname{Nuclear Physics A}

\runauth{D.V. Perepelitsa}

\jid{nupha}

\jnltitlelogo{Nuclear Physics A}

\usepackage{amssymb}

\usepackage{lineno}

\usepackage[figuresright]{rotating}

\begin{document}

\begin{frontmatter}



\dochead{}

\title{Hard processes in small systems}


\author[0]{Dennis V. Perepelitsa}

\address[0]{University of Colorado Boulder, 390 UCB, Boulder, CO 80309}

\begin{abstract}
These proceedings from the Quark Matter 2017 conference give an
overview of the latest experimental results on hard processes in small
collision systems at RHIC and the LHC, discuss their implications, and
consider several prospects for future measurements.
\end{abstract}


\end{frontmatter}


\section{Introduction}

These proceedings give an overview of high transverse momentum
($p_\mathrm{T}$) jet, hadron and electroweak probes of small
(e.g. proton--nucleus or $p$+A) collision systems. They are organized
around three components of such a collision: the initial state of the
nucleus, the initial state of the proton, and the final state of the
system. I emphasize results which are new since the previous Quark
Matter conference, attempt to place them into context with other
experimental data and theoretical ideas, and suggest avenues for
future work.

\section{Initial state of the nuclear wavefunction} 

Measurements of the inclusive production of jets and charged
particles, the most abundant QCD final-state objects, are a
fundamental way to characterize the partonic content of, and
fragmentation process in, a hadronic collision system. Some
measurements of these signatures in $5.02$~TeV $p$+Pb collision data
recorded in early 2013 at the LHC initially suggested that charged
particle production rates at large $p_\mathrm{T}$ and the
jet-to-particle fragmentation function may be modified to a
surprisingly large degree with respect to proton--proton ($pp$)
collisions. However, these measurements lacked directly measured $pp$
comparison data at the same energy and thus generally relied on
extrapolated references with large uncertainties.

\begin{figure*}[t]
\begin{center}
\includegraphics*[width=0.41\linewidth]{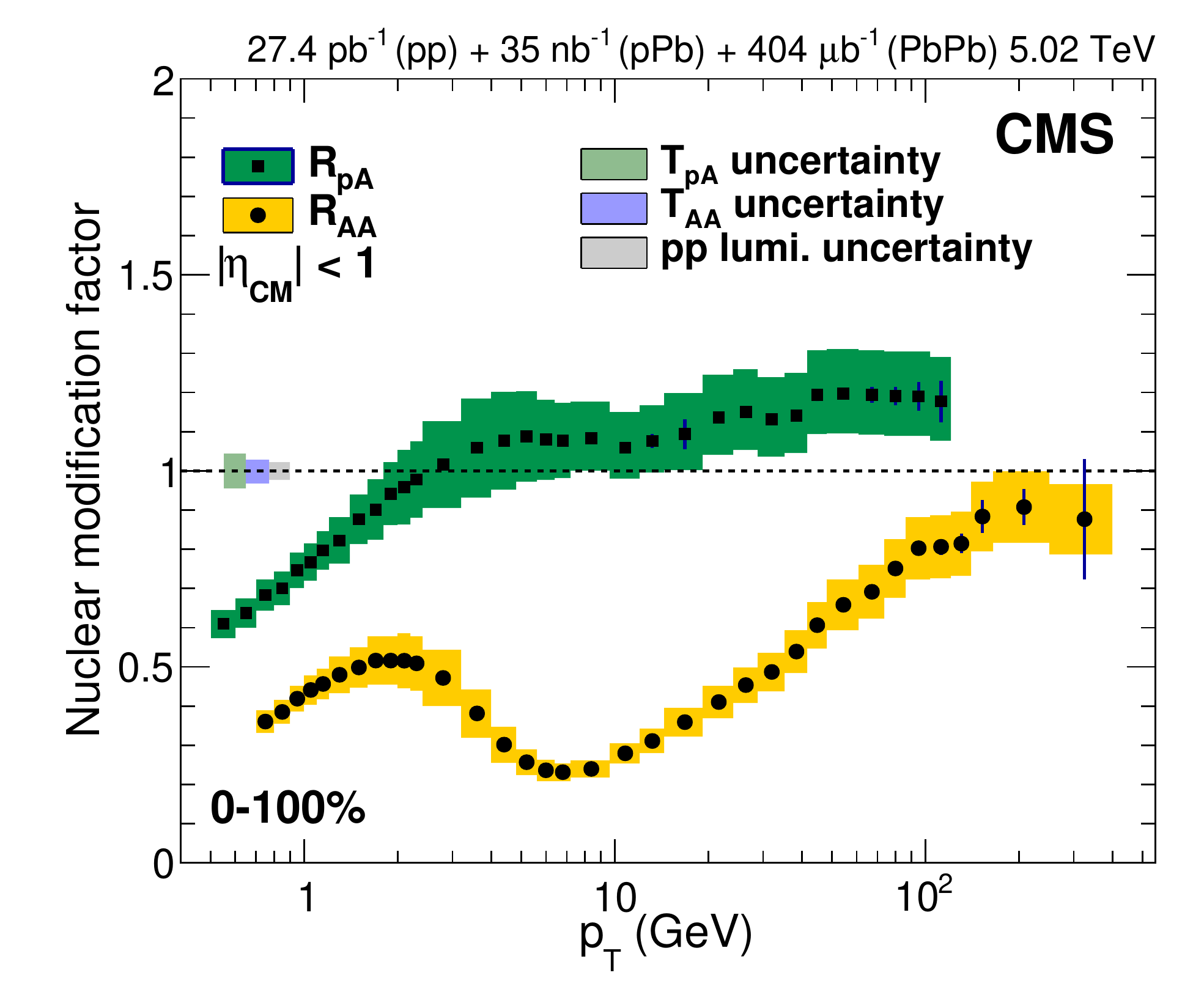}
\includegraphics*[width=0.58\linewidth]{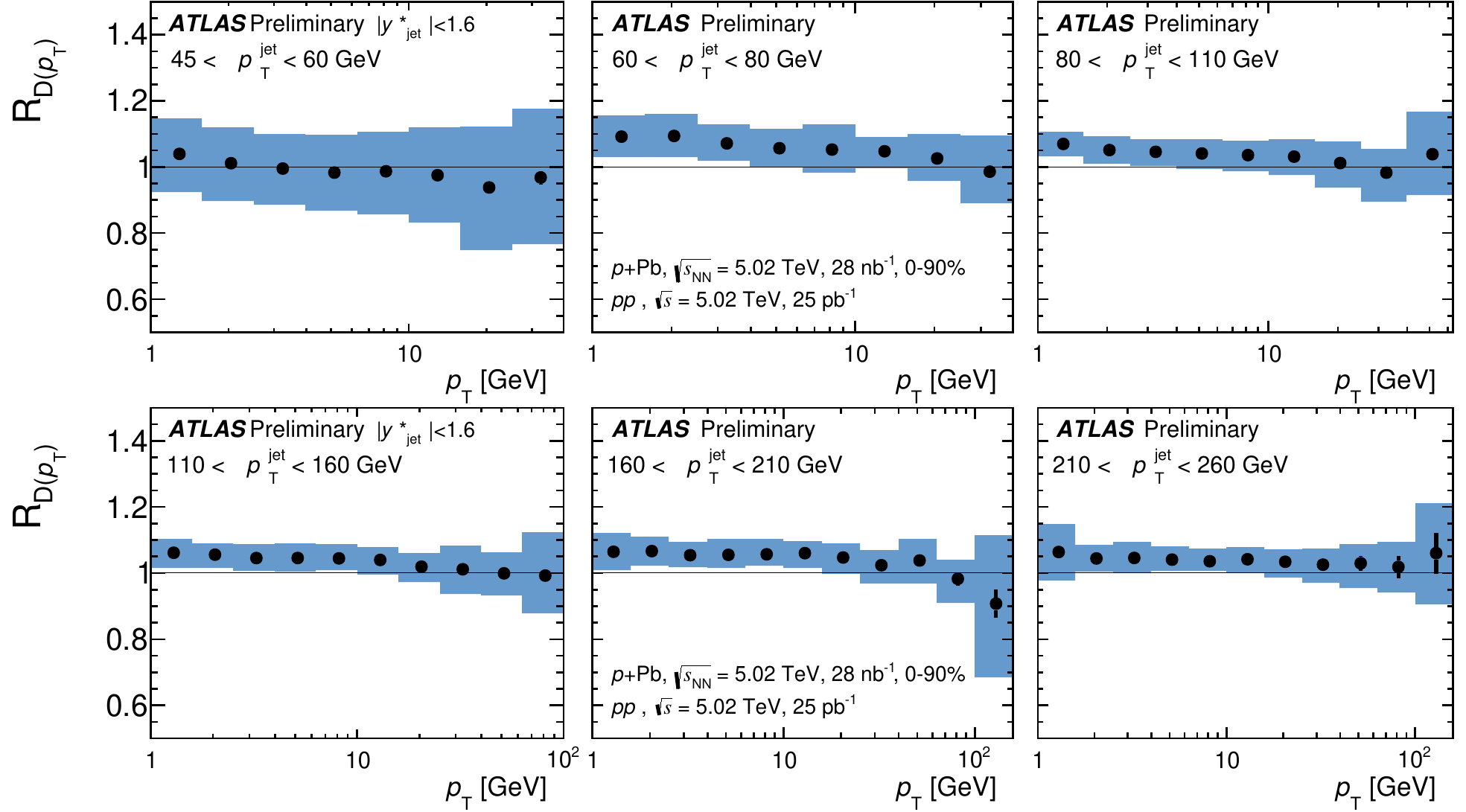}
\caption{Measurements of the nuclear modification factor for charged
  particles (left, from CMS~\cite{Khachatryan:2016odn}) and
  jet-to-particle fragmentation function (right, from
  ATLAS~\cite{ATLAS:2017mjv}) in $5.02$~TeV $p$+Pb collisions, each
  using a directly measured $5.02$~TeV $pp$ reference.
 \label{fig:fig1}}
\end{center}
\end{figure*}

At this conference, an updated set of experimental results were
presented which use a directly-measured $5.02$~TeV $pp$ reference from
the high-luminosity data recorded in late 2015 at the LHC. Two
important examples are shown in Fig.~\ref{fig:fig1}. Measurements of
charged particle production rates from ATLAS~\cite{ATLAS:2016kvp} and
CMS~\cite{Khachatryan:2016odn} indicate that they exhibit only a
modest enhancement in the anti-shadowing region, $p_\mathrm{T} \approx
40-80$~GeV and are otherwise consistent at lower $p_\mathrm{T}$ with
the initial measurement by ALICE~\cite{Abelev:2014dsa}. The
enhancement at large hadron-$p_\mathrm{T}$ is qualitatively similar to
that observed in the ATLAS measurement of inclusive jet
production~\cite{ATLAS:2014cpa}. Additionally, an updated measurement
of jet fragmentation functions from ATLAS~\cite{ATLAS:2017mjv} showed
that they are consistent with those in $pp$ data, now agreeing with a
preliminary CMS measurement~\cite{CMS:2015bfa}. Thus, the latest
available data on inclusive charged hadron production, jet production,
and jet-to-hadron fragmentation together give a mutually consistent
picture of the basic QCD hard scattering process in $p$+A collisions.

The inclusion of $5.02$~TeV $pp$ reference data has also brought
clarity to the interpretation of measurements in other channels. For
example, it has eliminated the theoretical uncertainties in previous
measurements by ATLAS of how $Z$ boson production rates are modified
in $p$+Pb collisions, demonstrating that they are consistent with
expectations from analyses of parton distribution functions in nuclei
(nPDFs)~\cite{ATLAS:2016ckg}. On the other hand, the inclusion of $pp$
data in measurements of dijet production by CMS has revealed that
current nuclear PDF sets are insufficient to describe modifications in
certain kinematic regions~\cite{CMS:2016kjd}. A particular
disagreement is found in the so called EMC region (large nuclear-$x
\gtrsim 0.2$) at large-$Q^{2}$. Measurements in this kinematic region
are particularly important for giving context to new results presented
at this conference by ATLAS on the suppression of TeV-scale jets in
Pb+Pb collisions~\cite{ATLAS:2017wvp}: the $p$+Pb measurements
constrain how much of the observed suppression in Pb+Pb may be
expected from initial state, rather than final state, effects. In any
case, the improved dijet and electroweak boson data are now being used
in the next generation of global nPDF extractions such as the EPPS16
analysis in Ref.~\cite{Eskola:2016oht}.

\subsection{Photo-nuclear processes and precision electroweak probes}

\begin{figure*}[t]
 \begin{minipage}[b]{0.39\linewidth}
\includegraphics*[width=\linewidth]{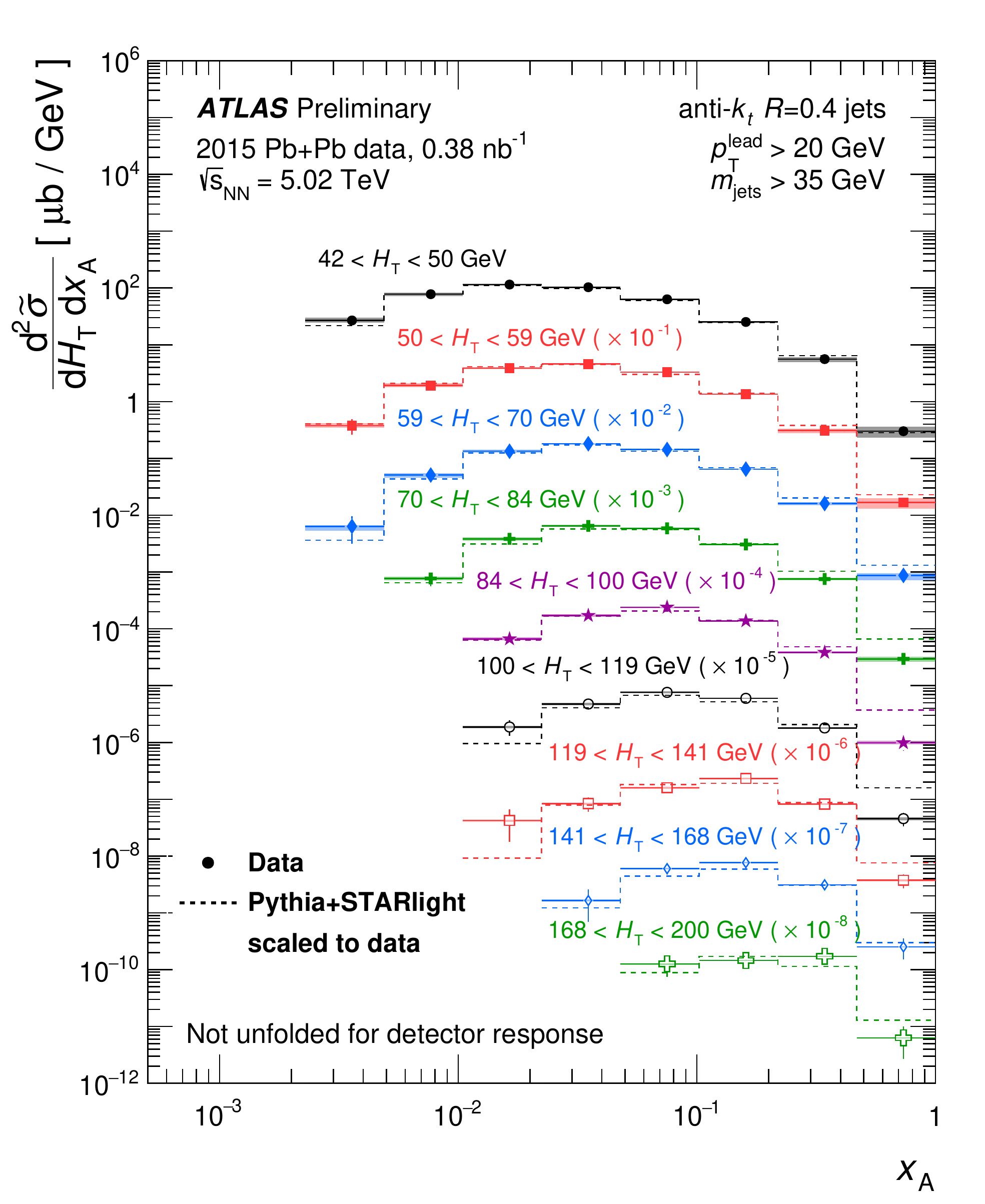}
  \end{minipage} 
  \begin{minipage}[b]{0.61\linewidth}
\includegraphics*[width=0.8\linewidth]{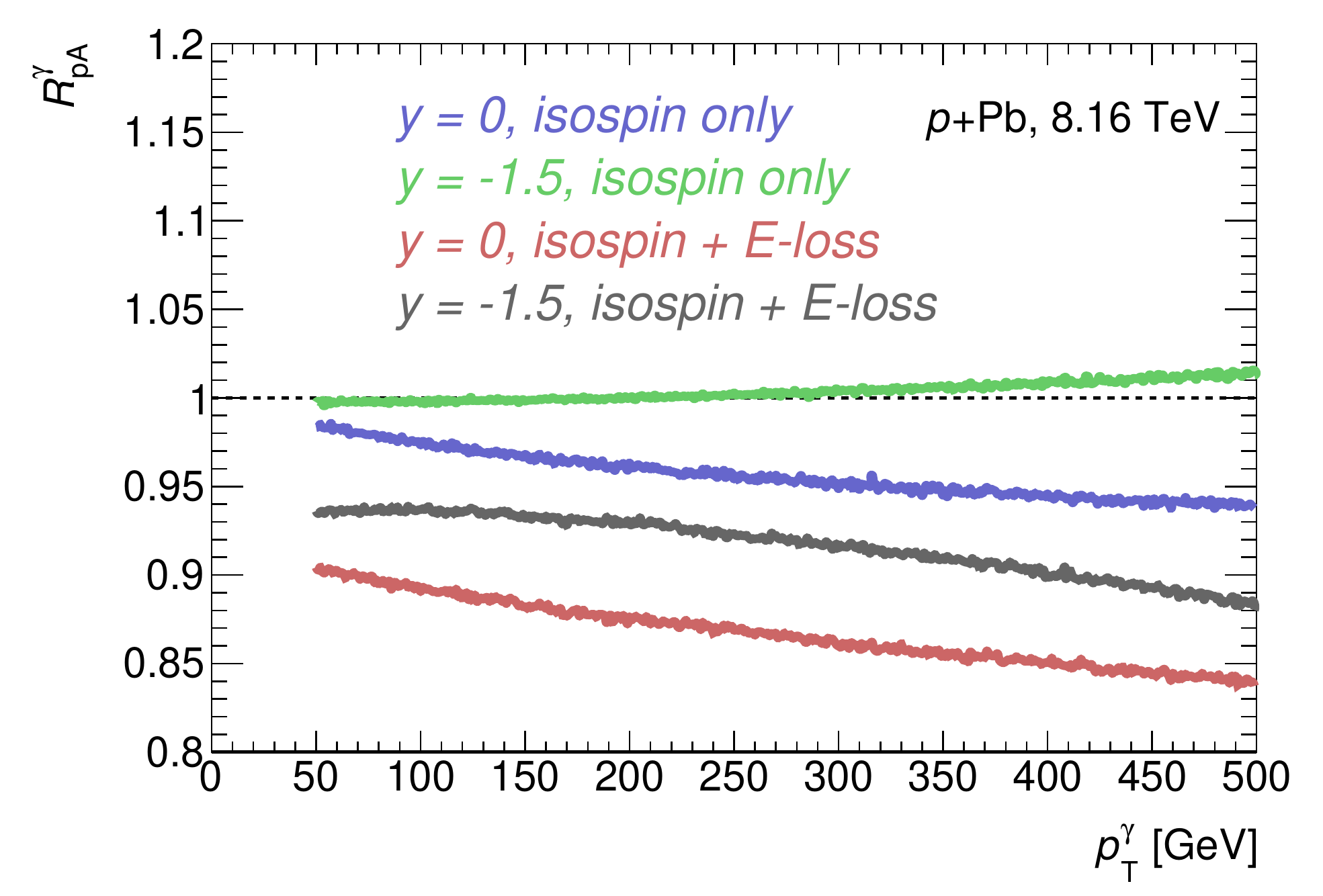}
\caption{{\em Left:} Measurement of photonuclear dijet production in
  $5.02$~TeV Pb+Pb collisions by ATLAS~\cite{ATLAS:2017kwa}. {\em
    Right:} Calculation of the role of initial state energy loss
  effects in inclusive direct photon production in $8.16$~TeV $p$+Pb
  collisions~\cite{Chien:2015vja,Vitev:2008vk}.
 \label{fig:fig2}}
  \end{minipage} 
\end{figure*}

Despite the breadth of these measurements, it is useful to further
improve the accessible kinematic range and experimental precision of
measured initial state effects. One promising possibility is the use
of hard photo-nuclear processes in ultra-peripheral nucleus-nucleus
collisions~\cite{Strikman:2005yv} (i.e. those with an impact parameter
much larger than the nuclear radii). Schematically, a quasi-real
photon emitted by one nucleus may split into a quark-antiquark dipole
and scatter with a gluon in the other nucleus, generically resulting
in dijet production. These collisions can be distinguished from
ordinary low-multiplicity inelastic (peripheral) Pb+Pb collisions
through their event topology, which features large pseudorapidity gaps
in particle production on one side of the detector. Thus these
processes allow for measurements of nuclear effects on jet production
without the substantial experimental complications introduced by the
high-multiplicity underlying event present in hadronic nuclear
collisions, allowing jets to be measured at low $p_\mathrm{T}$
values. In this conference, ATLAS presented a preliminary measurement
of photo-nuclear dijet production cross-sections over an expansive
kinematic range~\cite{ATLAS:2017kwa} (Fig.~\ref{fig:fig2}, left). In
particular, this measurement is sensitive to nuclear PDF effects in a
$(x_\mathrm{A}, Q^{2})$ range normally in accessible in existing
fixed-target deep inelastic scattering data and the LHC dijet and
electroweak data described above. When published in a fully unfolded
form, this measurement has the potential to provide a large input to
nPDF analyses in regions not directly constrained by any data. The CMS
and ALICE experiments presented new results on the photoproduction of
heavy vector mesons in ultra-peripheral Pb+Pb
collisions~\cite{UPC1,UPC2}, providing complementary data at lower
$Q^2$.

The $8.16$~TeV $p$+Pb data recently collected at the LHC in late 2016
also offers new opportunities for further precision studies of the
initial nuclear state. In particular, the substantially larger
luminosity of this dataset compared to the lower-energy $p$+Pb data
collected in 2013 ($170$ nb$^{-1}$ vs. $31$ nb$^{-1}$) as well as the
modestly larger cross-sections for hard-scattering processes will
enable high-statistics measurements in electroweak channels. Unlike
the situation at the time of the $5.02$~TeV $p$+Pb data-taking,
precision measurements of QCD and electroweak processes in the
existing high-statistics $8$~TeV $pp$ data collected in 2012 can be
used as the reference for $8.16$~TeV $p$+Pb collisions (with only a
modest correction for the small collision energy difference). To give
one example of the quality of the readily available reference data,
the systematic uncertainty in existing measurements of the photon
production cross-section in $8$~TeV $pp$ collisions~\cite{Aad:2016xcr}
reaches $\sim2$\% for $80$--$300$~GeV photons at mid-rapidity,
including luminosity uncertainties.

I give three specific examples of electroweak measurements enabled by
this large-luminosity dataset: (1) Inclusive direct photon production
rates can be measured at central and forward rapidities out to several
hundred GeV, where calculations suggest they will be sensitive to
non-trivial initial state energy loss
effects~\cite{Chien:2015vja,Vitev:2008vk} (Fig.~\ref{fig:fig2},
right). (2) Di-photon production can be measured for the first time in
a heavy ion context. This fundamental QCD process is sensitive,
through a box diagram with internal quark loop, to the gluon
distribution in both beams, allowing for experimentally clean access
to the low nuclear-$x$ regime. (The scientific motivation is similar
as in Ref.~\cite{Kovner:2015rna} but with a kinematic region closer to
that previously measured in the LHC as in Ref.~\cite{Aad:2011mh}.) (3)
Top quarks can be observed for the first time in heavy ion collisions,
for example through $t\bar{t}$ production in which both top quarks
produce a high-$p_\mathrm{T}$ electron or muon in their decay
chain. Such a measurement would be sensitive to nuclear modifications
of the gluon PDF at large nuclear-$x$, a kinematic region which is
otherwise difficult to access experimentally~\cite{dEnterria:2015mgr}.

\subsection{Measurements as a function of centrality}

In addition to measurements in minimum bias (inclusive in
event-activity) $p$+A collisions, the centrality- or
event-activity-dependence of hard process rates in small collision
systems is interesting. Such measurements are traditionally motivated
as a way to probe the impact parameter dependence of nPDF modification
or, for example, the nuclear thickness dependence of initial state
energy loss effects. More generally, they test the combined
understanding of soft and hard particle production processes in these
collision systems.

In $p$+A collisions, the first attempts to measure centrality-selected
hard process rates have generally found that they are inconsistent
with the binary-collision-scaled expectation from $pp$
collisions. Frequently, these observations are discussed as arising
from a ``bias'', for example, in the event selection or estimation of
geometric parameters. In fact, there are at least two distinct
effects, with distinct physics origins, signs, and kinematic
dependences. The first effect is the conventional {\em centrality
  bias}, most visible at moderate-$p_\mathrm{T}$ values
($\lesssim10$~GeV or $\lesssim50$~GeV at RHIC and LHC energies,
respectively), which causes a kinematics-independent increase
(decrease) of the yield in central (peripheral) collisions. This
effect arises because the increased multiplicity associated with the
presence of a hard parton--parton scattering causes hard-scatter $p$+A
events to be categorized in a higher-activity centrality class than
the centrality calibration, derived from analyzing minimum-bias events
without a hard scatter, would predict.

\begin{figure*}[t]
\begin{center}
\includegraphics*[width=0.51\linewidth]{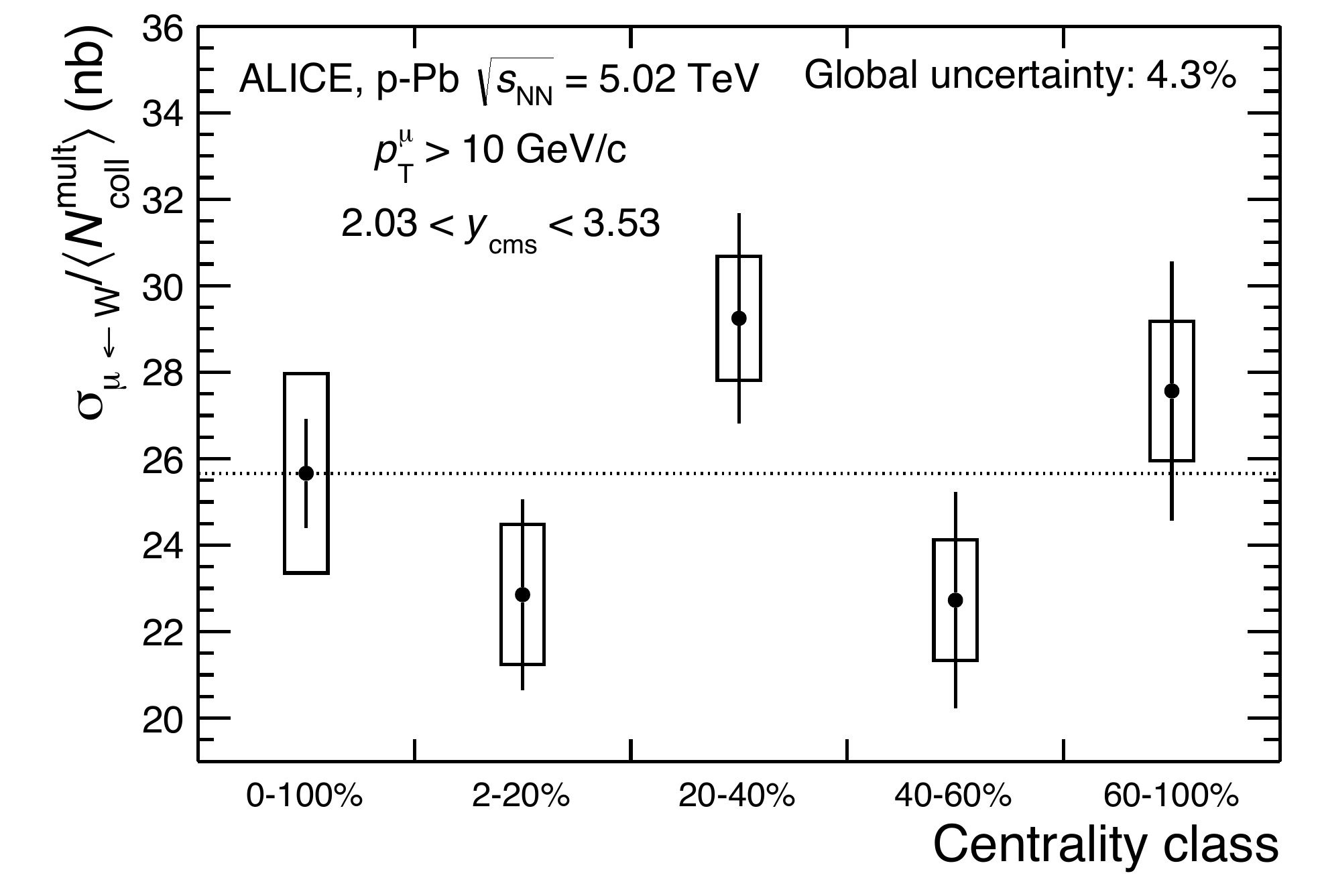}
\includegraphics*[width=0.47\linewidth]{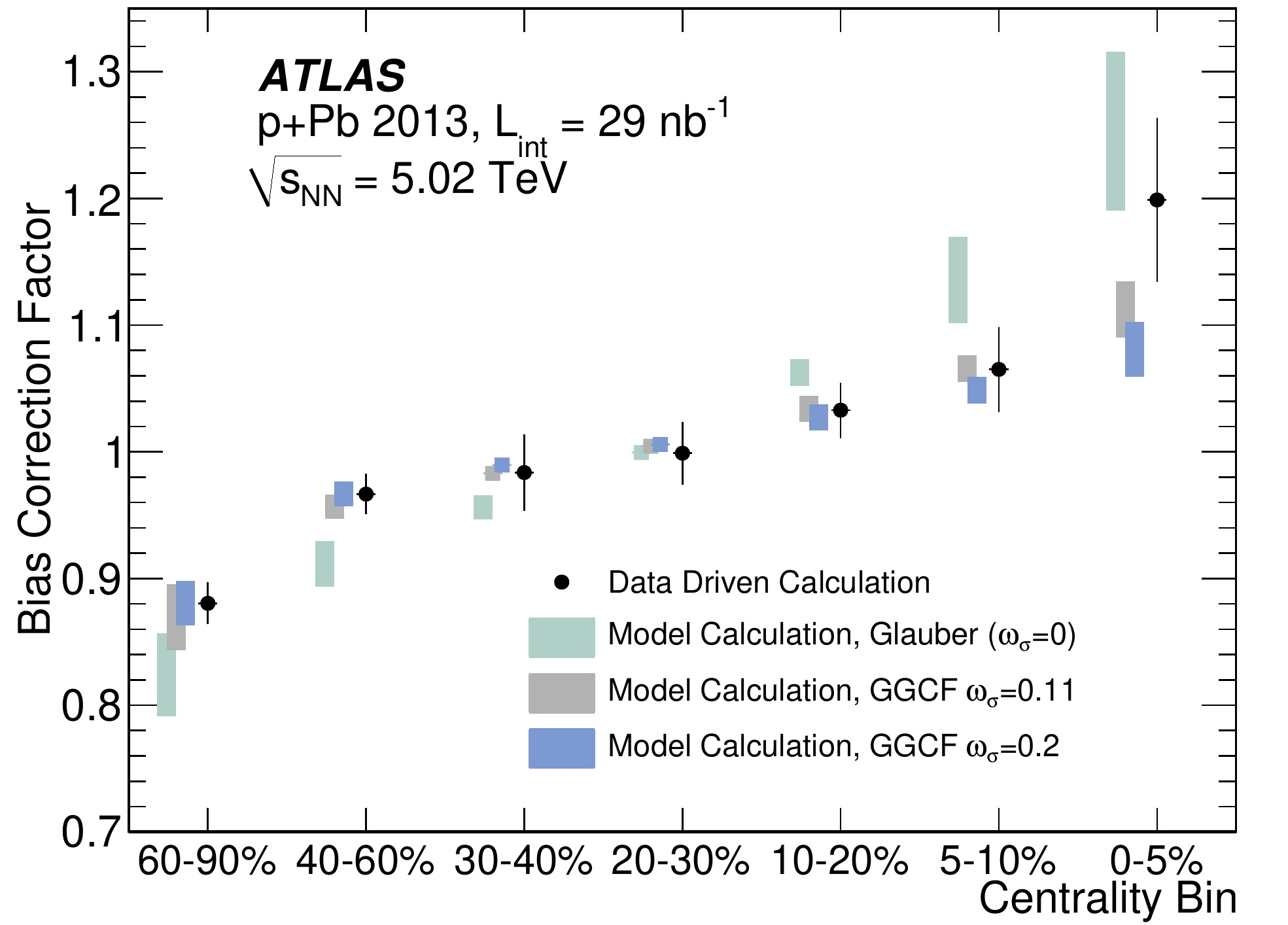}
\caption{{\em Left:} ALICE measurement of the centrality-dependence of
  $W^\pm$ production in $5.02$~TeV $p$+Pb
  collisions~\cite{Alice:2016wka} within the hybrid
  model~\cite{Adam:2014qja}. {\em Right:} Calculation of centrality
  bias factors with data-driven~\cite{Aad:2015gta} and
  analytic~\cite{Perepelitsa:2014yta} methods for $p$+Pb collisions in
  ATLAS.
 \label{fig:fig3}}
\end{center}
\end{figure*}

The experiments have found several methods to correct for or eliminate
this first bias. I summarize them here: (1) In the ``hybrid method''
developed by ALICE, $p$+Pb events are categorized by their energy
signature in the zero-degree calorimeter far downstream of the Pb
nucleus~\cite{Adam:2014qja}, which has a large pseudorapidity
separation from the mid-rapidity region. Since the distribution of
energies in this kinematic region is difficult to associate to a
Glauber model, the geometric parameters are estimated by assuming that
they scale with the soft particle multiplicity elsewhere in the
event. Using this method, ALICE has presented a new measurement of
$W^{\pm}$ production which demonstrates a binary-collision
scaling~\cite{Alice:2016wka} (Fig.~\ref{fig:fig3}, left). (2) One may
calculate the quantitative effects of the bias on measured yields
based on the correlation between hard process rates and underlying
event activity in $pp$ collisions, such as the method used by
ATLAS~\cite{Perepelitsa:2014yta} which builds on previous work in this
vein by PHENIX~\cite{Adare:2013nff} (Fig.~\ref{fig:fig3}, right). When
applied to $p$+Pb collision data, these corrections generally restore
the expected binary collision scaling for electroweak boson and
quarkonia production, as shown by ATLAS~\cite{Aad:2015gta}. (3) Third,
one may argue that the bias arises from an incomplete modeling of
$p$+A collision geometries. In particular, event-to-event color
fluctuations in the configuration of the proton, which in the context
of geometric modeling is implemented in the so-called Glauber-Gribov
Color Fluctuation (GGCF) model~\cite{Aad:2015zza,Alvioli:2013vk}, may
play an important role. In fact, some measurements, such as that of
femtoscopic radii in centrality-selected $p$+Pb collisions presented
by ATLAS at this conference~\cite{Aaboud:2017xpw}, suggest a
preference for the broader $N_\mathrm{part}$ distribution in the GGCF
model over that in the traditional Glauber model. When geometric
parameters are estimated under the GGCF model, hard-process yields
(such as $Z$ bosons in Ref.~\cite{Aad:2015gta}) are closer to the
$pp$-based expectation before any additional corrections.

The second ``bias'' is a physics effect which for mid-rapidity
processes is relevant only at large $p_\mathrm{T}$ ($>10$~GeV and
$\gtrsim200$~GeV at RHIC and the LHC, respectively). Although it may
manifest as a bias in observables such as nuclear modification
factors, it has a distinctive experimental signature from the
conventional centrality bias described above. Here I briefly describe
the essential features observed in data: (1) Rates of
high-$p_\mathrm{T}$ jet production at mid-rapidity at RHIC are
observed to be ``split'' in centrality: they are suppressed (enhanced)
in central (peripheral) collisions, in a way that grows with
increasing jet-$p_\mathrm{T}$, but in such a way that they integrate
to a null effect for minimum bias collisions~\cite{Adare:2015gla}. (2)
Rates of high-$p_\mathrm{T}$ jet production at the LHC are modified in
an analogous way at a much larger jet-$p_\mathrm{T}$ range but, when
measuring at mid-rapidity, in a similar $x_\mathrm{T} = 2 p_\mathrm{T}
/ \sqrt{s}$ range, $x_\mathrm{T} \gtrsim
0.1$~\cite{ATLAS:2014cpa}. (3) At systematically more proton-going
rapidities, the centrality-splitting effect is evident at
systematically smaller jet-$p_\mathrm{T}$ values and nuclear
modification factors scale with $p_\mathrm{T} \cosh(y) \sim
E_\mathrm{jet}$ across six units of rapidity. In a leading order
picture, jets with total energy $E_\mathrm{jet}$ in the forward region
are most commonly produced from parton--parton configurations with
Bjorken-$x_p$ in the proton given by $x_p \approx 2 E_\mathrm{jet} /
\sqrt{s}$ (as noted for example in Ref.~\cite{Alvioli:2014eda}). (4)
Measurements of the pseudorapidity distributions for dijet pairs shows
that their production at forward rapidities is systematically
suppressed (enhanced) with selections on large (small) event activity
~\cite{Chatrchyan:2014hqa}. (5) Finally, a control analysis of how
transverse energy production at large (one-sided) rapidity broadly
depends on parton--parton scattering kinematics~\cite{Aad:2015ziq}
refutes the suggestion that this effect is caused by a
rapidity-separated ``energy conservation'', extending a similar
conclusion reached by studying hard processes at mid-rapidity in
Ref.~\cite{Adare:2013nff}.

Together, the collision-energy- and rapidity-dependence of the
centrality-splitting effect in data strongly suggests that it is
controlled predominantly by only one kinematic variable: the
Bjorken-$x_p$ of the hard-scattered parton in the proton beam or,
perhaps, Feynman-$x_\mathrm{F} = x_p - x_\mathrm{A}$ which is
numerically similar to $x_{p}$ in the forward region. As will be
described below, this observation suggests that the data is sensitive
to an $x_p$-dependent property of the proton wavefunction before the
collision. However, an initial state energy loss effect which depends
predominantly on $x_\mathrm{F}$~\cite{Kang:2015mta} may also play some
role.

\section{Initial state of the proton wavefunction}

One efficient explanation of the observed centrality-dependence of
hard process rates at large $x_p > 0.1$ is that it arises from a
change in the properties of protons with a large-$x_p$ parton. This
and similar ideas are developed in
Refs.~\cite{Alvioli:2014eda,Bzdak:2014rca,Armesto:2015kwa,Kordell:2016njg}. Schematically,
the {\em shrinking proton} picture supposes that protons in a
large-$x_p$ configuration have a smaller transverse size, fewer
additional partons, and interact with other nucleons with a smaller
cross-section, than average protons. When protons in a large-$x_p$
configuration pass through the nucleus, they interact with fewer
nucleons than a proton in an average configuration would, and
therefore a smaller centrality signature is produced in the collision
than is expected given its underlying geometric configuration. Thus
production rates in minimum bias (e.g. centrality-averaged) collisions
are unmodified, but events with a large-$x_p$ proton are
systematically redistributed to a more peripheral
classification. Since the size or interaction strength of the proton
decreases with increasing $x_p$, this provides an efficient
explanation for the observed set of data at multiple rapidities and
collision energies.

\nocite{McGlinchey:2016ssj,PHENIXOverview,Blok:2012jr}

\begin{figure*}[t]
\begin{center}
\includegraphics*[width=0.66\linewidth]{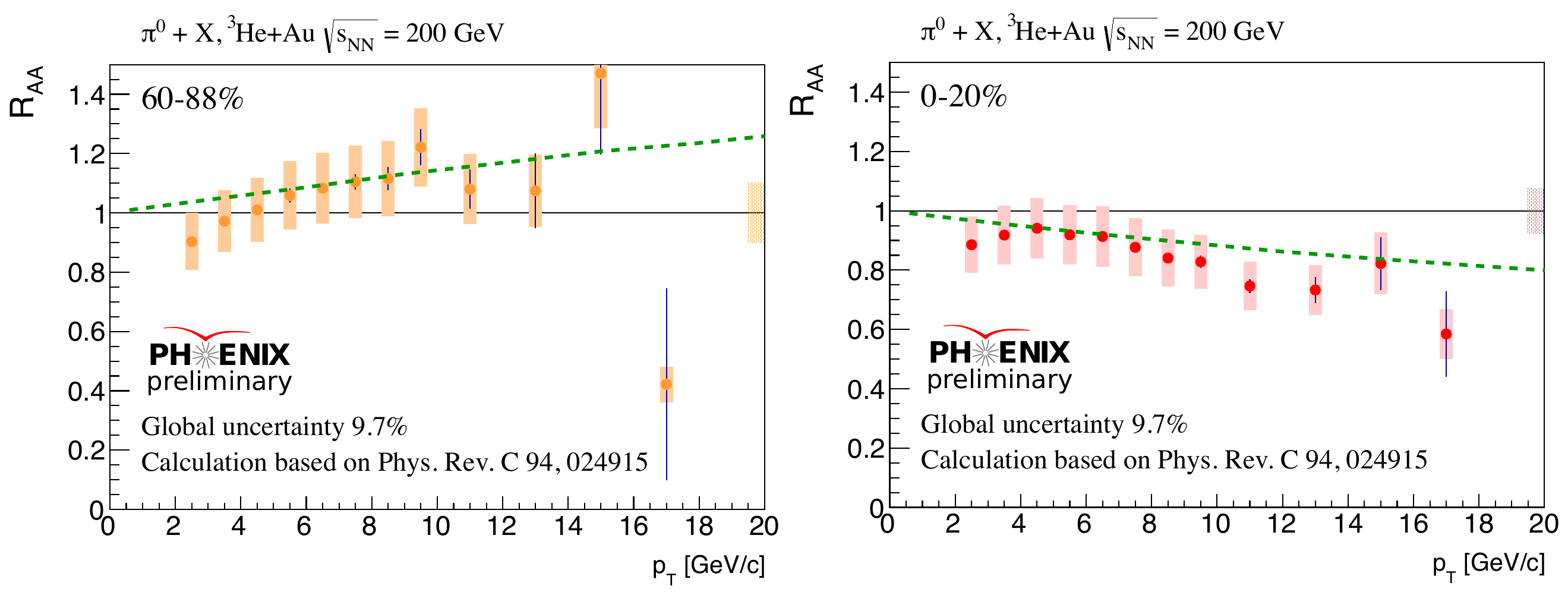}
\includegraphics*[width=0.33\linewidth]{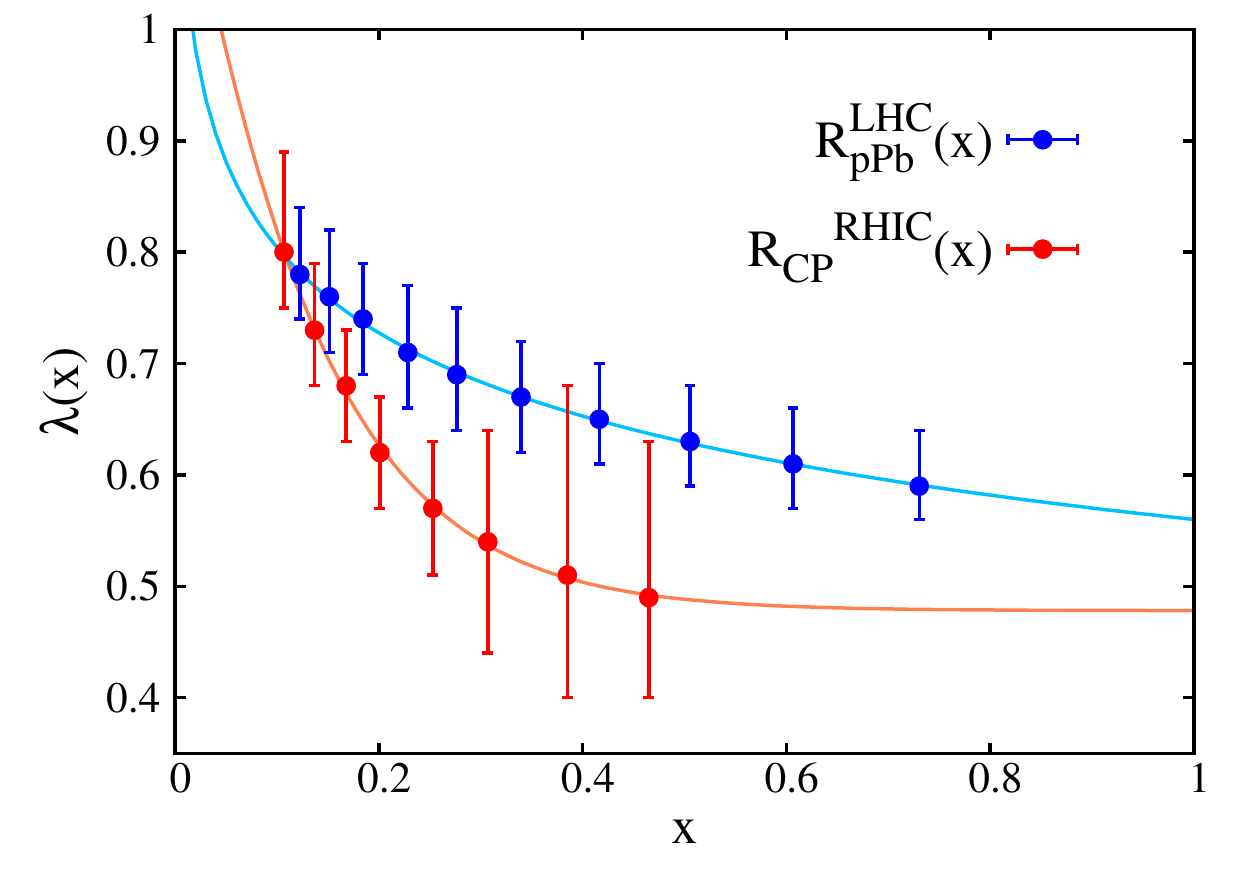}
\caption{{\em Left and center:} Measurement by PHENIX of the
  ``centrality splitting'' in high-$p_\mathrm{T}$ $\pi^0$ production
  in $^{3}$He+Au collisions~\cite{PHENIXOverview} compared to the
  calculation in Ref.~\cite{McGlinchey:2016ssj}. {\em Right:}
  Extracted values of $\lambda(x_p)$, the ratio of the interaction
  cross-section for nucleons in a large-$x_p$ configuration to that
  for average configurations, at RHIC (red) and LHC (blue)
  energies~\cite{PerepelitsaStrikman}.
 \label{fig:fig4}}
\end{center}
\end{figure*}

I list several ways in which the shrinking proton picture can be
further tested experimentally or additional ways in which aspects of
this physics can be explored: (1) A unique test may be performed with
the recently collected $200$~GeV $p$+Au and $^{3}$He+Au collision data
recorded at RHIC. The centrality-splitting effect is expected to be
larger in the $p$+Au system relative to that in the $d$+Au system
since the absence of a (non-shrinking) nucleon in the projectile beam
will strengthen the relative effect, and vice versa for the
$^{3}$He+Au system~\cite{McGlinchey:2016ssj}. Preliminary data
presented by PHENIX on $\pi^0$ production in $^{3}$He+Au
collisions~\cite{PHENIXOverview} follows this expected
projectile-species dependence (Fig.~\ref{fig:fig4}, left). An
analogous set of preliminary $p$+Au data observes a large
central-to-peripheral difference as expected, but this difference
modulates an overall suppression across all centrality selections
(including minimum-bias collisions), suggesting some impact from the
$pp$ reference common to each $p$+Au centrality selection which should
be further explored. (2) In addition to the overall transverse size of
the beam-remnant, the momentum and spatial structure of the remainder
of the proton in a large-$x_p$ configuration could be studied. One
experimental prospect enabled by the high-luminosity $p$+Pb data is
the study of so-called double-parton scattering events, in which two
distinct partons in the proton participate in independent
hard-scatterings with partons in the
nucleus~\cite{Blok:2012jr}. Schematically, after first requiring
evidence of a large-$x_p$ scattering, one could explore the how the
kinematics of the second, independent hard-scattering are correlated
with event activity. (3) The RHIC and the LHC data could be analyzed
within a consistent theoretical framework to extract how the proton
size shrinks with $x_p$ at different collision energies. Such an
analysis is sensitive to how the cross-section for small-proton
configurations grows with energy relative to that for average
configurations (with the preliminary result of such an
extraction~\cite{PerepelitsaStrikman} shown in Fig.~\ref{fig:fig4},
right). (4) Measurements of the centrality-dependence of other (for
example, electroweak) processes at large-$x_p$ can test whether the
shrinking of the proton depends on the flavor of the parton, and
otherwise confirm that the shrinking of the proton in the initial
state is process-independent.

\section{Final state effects in small systems}

\begin{figure*}[t]
\begin{center}
\includegraphics*[width=0.47\linewidth]{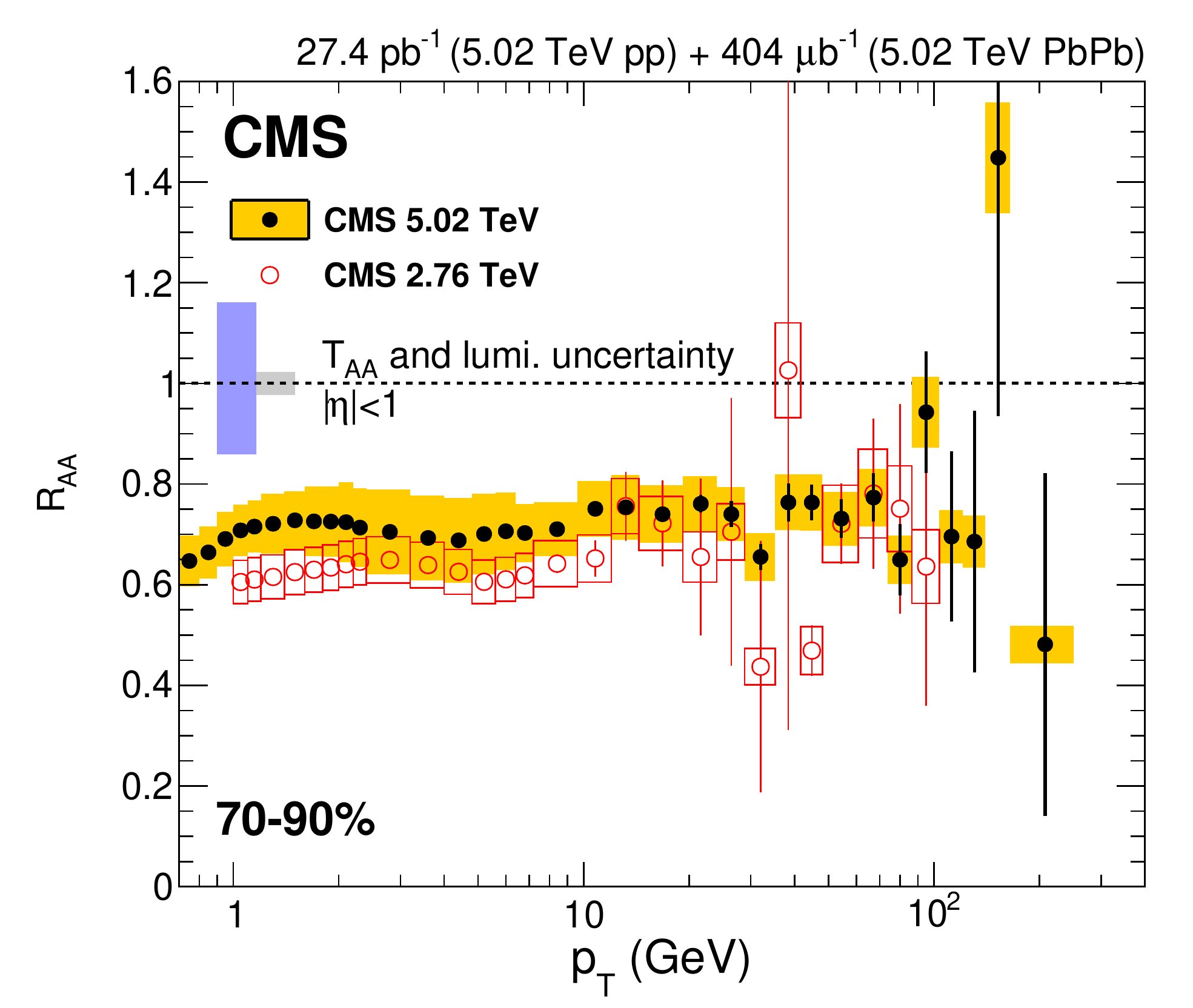}
\includegraphics*[width=0.42\linewidth]{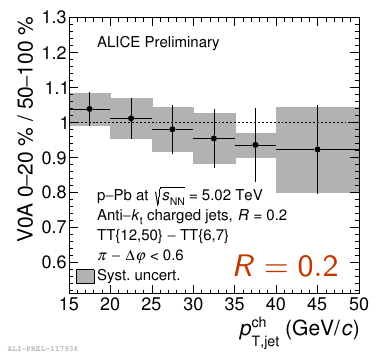}
\caption{{\em Left:} CMS measurement of the $R_\mathrm{AA}$ for
  charged particles in 70--90\% central
  ($\smash{\left<N_\mathrm{part}\right>} \approx 11 $) $5.02$~TeV
  Pb+Pb collisions~\cite{Khachatryan:2016odn}. {\em Right:} ALICE
  measurement of the hadron--triggered recoil jet $p_\mathrm{T}$
  spectrum, ratio of 0--20\% $p$+Pb events to that in 50--100\%
  events~\cite{ALICEOverview}. \label{fig:fig5}}
\end{center}
\end{figure*}

In addition to the initial state physics motivations described above,
hard processes may also be used to test for the final-state energy
loss of partons traversing a small-sized region of dense colored
matter or quark-gluon plasma which, as is recently suggested by
experimental signatures in the soft sector, may be created in these
collision systems. At this conference, new results were presented on
the suppression of inclusive jet and hadron spectra in peripheral
nucleus--nucleus (A+A) collisions, which also feature a small nuclear
overlap. For example, a new measurement of the hadron $R_\mathrm{AA}$
by CMS reveals that is suppressed by $30$\% in 70-90\% Pb+Pb
collisions (Fig.~\ref{fig:fig5}, left). Pb+Pb events in this
centrality range include eleven participating nucleons on average,
only slightly more than the average number in $p$+Pb
collisions. Furthermore, a simple comparison of the transverse
geometry of very peripheral A+A and central $p$+A collisions in a
Glauber or state-of-the-art hydrodynamic model~\cite{Weller:2017tsr}
reveals that they are not as different as one may (naively)
expect. While there are plausible reasons that $p$+A collisions may
behave differently than A+A collisions with a similar transverse
geometric extent, the absence of obvious jet quenching signatures in
$p$+A collisions is notable given the large suppression easily
observed in even very glancing A+A collisions.

Given the complications in interpreting the centrality-selected hard
process measurements described above (some trivial, some due to
non-trivial but ultimately unrelated large-$x$ physics), parton energy
loss effects may instead be probed event-by-event through intra-event
momentum correlations which are expected to be modified in the
presence of energy loss. One example is the azimuthal anisotropy $v_2$
of high-$p_\mathrm{T}$ charged particles in high-activity events. In
fact, a previous measurement by ATLAS observed that the
rapidity-separated azimuthal correlation function for $p_\mathrm{T} =
10$~GeV particles exhibits a visible near-side peak, with $v_{2}
\approx 0.05$, in the most central $1$\% of $5.02$~TeV $p$+Pb
events~\cite{Aad:2014lta}. In A+A collisions, this signature would be
traditionally interpreted as a path-length-dependent modification in
particle production resulting from final-state energy loss. Another
example is the $p_\mathrm{T}$ balance between high-$p_\mathrm{T}$
final state objects, which does not rely on an unbiased estimate of
geometric quantities. At this conference, a preliminary measurement of
the $p_\mathrm{T}$ spectrum of high-$p_\mathrm{T}$ hadron-triggered
recoil jets was presented by ALICE~\cite{ALICEOverview}, concluding
that is is unmodified in 0--20\% central $p$+Pb collisions relative to
that in peripheral collisions (Fig.~\ref{fig:fig5}, right), placing a
limit on the possible amount of energy loss. Considered together, the
two results provoke the question of whether an onset of jet quenching
may be observed at some centrality interval between these two.

To resolve any potential tension between peripheral A+A and central
$p$+A data, I suggest efforts along three lines: (1) The experimental
selection and modeling of very peripheral A+A collisions, frequently
overlooked due to the low hard process yield and small overlap region,
should be re-examined. Uncertainties on geometric parameters are
significant in the most peripheral A+A collisions ($>15$\% in the CMS
$R_\mathrm{AA}$ measurement~\cite{Khachatryan:2016odn}), largely
driven by the uncertainty in the fraction of inelastic A+A collisions
selected by triggers. Although current measurements of the
$N_\mathrm{coll}$-scaling of electroweak bosons in Pb+Pb
collisions~\cite{ATLAS:2017zkv} offers some confirmation of the
geometric picture, they are statistically limited in peripheral
collisions. A fresh look, using the improved understanding of
contaminating EM processes and pseudorapidity-gap based event
classification, would be welcome. (2) Jet quenching calculations or
Monte Carlo models which correctly describe the full
centrality-dependence of jet quenching signatures in A+A collisions
should be challenged to simultaneously describe the available data in
very peripheral A+A and central $p$+A collisions. (3) Finally, the
signatures above can be explored in more detail in the high-luminosity
$8.16$~TeV $p$+Pb data. Using high-multiplicity triggers enabled
during data-taking, experiments can extend measurements of
two-particle correlations to higher $p_\mathrm{T}$ and a wider
centrality range. The data will also provide ample statistics for
jet+X $p_\mathrm{T}$-correlations: for example, results on photon--jet
correlations presented at this
conference~\cite{ATLAS:2016tor,CMS:2013oua} demonstrate that they can
be measured with good systematic control in $pp$ events. Based on
existing measurements of isolated photon production cross-sections in
$8$~TeV $pp$ collisions~\cite{Aad:2016xcr}, it is expected that the
$8.16$~TeV $p$+Pb data contains more than ten thousand photons with
$p_\mathrm{T}^{\gamma} > 40$~GeV in the 1\% highest-multiplicity
events, allowing for a high-statistics, well-controlled search for
energy loss effects.

Such questions could also be explored through collisions of
intermediate-sized nuclei. For example, a program of Ar+Ar collisions
at the LHC would achieve a large luminosity with a level of underlying
event activity amenable to precision measurements of energy loss
signatures even for low-$p_\mathrm{T}$ jets.

\section{Conclusion}

These proceedings discuss new results on hard processes in small
systems presented at the Quark Matter conference, focusing on the
insight they provide into the initial state of the nuclear and proton
wavefunctions and into any final state effects in the resulting
system. For each topic, I have attempted to suggest directions for
future measurements or analysis, which I hope will be available in
time for the next Quark Matter.





\bibliographystyle{elsarticle-num} \bibliography{dvp-QM17-proceedings}







\end{document}